\pdfoutput=1
\documentclass[conference,letterpaper]{IEEEtran}

\usepackage[T1]{fontenc}
\usepackage[utf8]{inputenc}
\input glyphtounicode
\usepackage{amsmath,graphicx,color,cite}
\graphicspath{{./figures/}}
\usepackage{cite}
\usepackage{flushend}
\usepackage{hyperref}

\newcommand{\vect}[1]{\boldsymbol{#1}}
\newcommand*\concat{\mathbin{\|}}

\newcommand*\Figure[1]{Figure~\ref{#1}}
\newcommand*\Table[1]{Table~\ref{#1}}
\newcommand*\Section[1]{Section~\nameref{#1}}

\title{Simple Signal Extension Method for\\ Discrete Wavelet Transform}
\author{
	\IEEEauthorblockN{David Barina, \qquad Pavel Zemcik, \qquad Michal Kula}
	\IEEEauthorblockA{
		Brno University of Technology \\
		Czech Republic \\
		\{ibarina,zemcik,ikula\}@fit.vutbr.cz
	}
}

\begin{document}

\maketitle

\begin{abstract}
Discrete wavelet transform of finite-length signals must necessarily handle the signal boundaries.
The state-of-the-art approaches treat such boundaries in a complicated and inflexible way, using special prolog or epilog phases.
This holds true in particular for images decomposed into a number of scales, exemplary in JPEG 2000 coding system.
In this paper, the state-of-the-art approaches are extended to perform the treatment using a compact streaming core, possibly in multi-scale fashion.
We present the core focused on CDF 5/3 wavelet and the symmetric border extension method, both employed in the JPEG 2000.
As a result of our work, every input sample is visited only once, while the results are produced immediately, i.e. without buffering.
\end{abstract}


\section{Introduction}
\label{sec:introduction}

The discrete wavelet transform (DWT) is the signal-processing transform suitable as a basis for sophisticated compression algorithms.
Particularly, JPEG 2000 is an image coding system based on such compression technique.
The transform results in two subbands, corresponding to two FIR filters (a low-pass and high-pass ones).
With focus on computing demands, the transform is most commonly calculated using the lifting scheme.
This scheme reduces the number of arithmetic operations as compared with the original filter bank.
Considering finite signals, some special treatment of signal boundaries is unavoidable in both cases.
This induces special prolog and epilog phases, which cause delays at the beginning and end of the processing.
Such scenario breaks the ability for a continuous stream processing, especially considering the multi-scale decomposition.

In this paper, we propose a solution consisting of a linear mapping of input coefficients onto output ones.
This mapping is built above the lifting scheme.
Moreover, the mapping is referred to as the core in the rest of the paper.
The key idea is exploiting of the possibility of appropriate adjustments of the core, when it processes close to the signal boundaries.
For this reason, we speak about the mutable core.
The solution is demonstrated using the CDF 5/3 wavelet (used e.g. in JPEG 2000).

The rest of the paper is organized as follows.
\Section{sec:related-work} reviews the state of the art, especially the lifting scheme.
\Section{sec:proposed-method} proposes the core and the treatment of signal boundaries.
The purpose of \Section{sec:discussion} is to state our interpretations and opinions.
Finally, \Section{sec:conclusions} summarizes the work.

\section{Related Work}
\label{sec:related-work}

The discrete wavelet transform has undergone gradual development in the last few decades.
As a key advance in image processing, Cohen--Daubechies--Feauveau (CDF) \cite{Cohen1992} biorthogonal wavelets provided several families of symmetric wavelet bases.
Afterwards, W. Sweldens \cite{Sweldens1995,Sweldens1996,Daubechies1998} presented the lifting scheme which sped up such decomposition.
Following his work, any discrete wavelet transform can be decomposed into a sequence of simple filtering steps (lifting steps).
Readers not closely familiar to the DWT are referred to the excellent book \cite{Mallat2009} by S. Mallat.

The polyphase matrix \cite{Strang1996,Daubechies1998,Vetterli2014} is a convenient tool to express the transform filters.
The lifting scheme factorizes this matrix into a series of shorter filterings.
In detail, the polyphase matrix can be factorized \cite{Blahut2010}, so that
\begin{align}
	{P}(z) = \prod_{i=0}^{L-1}
	\left\{
	\begin{bmatrix}
		1 & S_i(z) \\
		0 & 1
	\end{bmatrix}
	\begin{bmatrix}
		1 & 0 \\
		T_i(z) & 1
	\end{bmatrix}
	\right\}
	\begin{bmatrix}
		K & 0 \\
		0 & 1/K
	\end{bmatrix},
\end{align}
where $K$ is a non-zero constant, and polynomials $S_i(z), T_i(z)$ represent the individual lifting steps.
Focusing on the CDF 5/3 wavelet as an example, the forward transform in can be expressed \cite{Daubechies1998} by the dual polyphase matrix
\begin{align}
	\label{eqn:polyphase-cdf53}
	\begin{split}
		\tilde{P}(z) & =
		\begin{bmatrix}
			1 & \alpha \left( 1 + z^{-1} \right) \\
			0 & 1
		\end{bmatrix}
		\begin{bmatrix}
			1 & 0 \\
			\beta \left( 1 + z \right) & 1
		\end{bmatrix}
		\begin{bmatrix}
			\zeta & 0 \\
			0 & 1/\zeta
		\end{bmatrix}
	\end{split},
\end{align}
where $\alpha,\beta$ are real constants, and the $\zeta$ is called the scaling factor.
For details about the lifting scheme, see \cite{Daubechies1998,Sweldens1996}.
Unlike a convolution scheme, the lifting allows \cite{Calderbank1998} formulation of the transforms mapping integers to integers.

To keep the total number of wavelet coefficients equal to the number of input samples, symmetric border extension is widely used.
The symmetric border extension method assumes that the input image can be recovered outside its original area by symmetric replication of boundary values.
A special variant (periodic symmetric extension) of this extension is employed in JPEG 2000 standard.

The treatment of the symmetric (or any other) border extension requires special prolog and epilog phases.
Considering the contemporary solutions, these phases cause increased delays at the beginning and end of the data processing.
Consequently, these phases break the stream processing.
Although this issue seems insignificant in single-level transform, it becomes a big problem considering the multi-scale processing.

The paper is mainly focused on contemporary processors equipped with some sort of CPU cache.
Inherently, the data processing should be performed in a single loop.
Excellent introduction to this topic can be found in \cite{Drepper2007}.

Originally, the problem of efficient implementation of the \mbox{1-D} lifting scheme was addressed in \cite{Chrysafis2000} by Ch. Chrysafis and A. Ortega.
Their general approach transforms the input data in a single loop.
Nonetheless, this is essentially the same method as the on-line or pipelined method mentioned in other papers (although not necessarily using lifting scheme nor \mbox{1-D} transform).
The key idea is to make the lifting scheme causal, so that it may be evaluated as a running scheme without buffering of the whole signal.
However, due to borders of the finite signals, some buffering is necessary at least at the beginning and end of the single-loop processing.
For \mbox{2-D} signals, the same idea was also discovered in another papers under various names, e.g. in \cite{Chaver2002,Shahbahrami2007,Shahbahrami2011,Chrysafis2000b,Oliver2005,Barina2015a}. 
The most sophisticated techniques, that we are aware of, were investigated by R. Kutil in \cite{Kutil2006}.
The author emphasizes that the main issue are the arduous prolog and epilog phases.
Many authors \cite{Meerwald2002,Tenllado2003,Chatterjee2002,Tao2008} also discovered a similar technique with a slightly different goal (a better utilization of cache lines).
This technique is referred to as the aggregation, strip-mine, or loop tiling.

Since this work is based on our previous works in \cite{Barina2015c,Barina2016,Barina2016b}, it should be explained what the difference between this work and the referenced papers is.
In \cite{Barina2015c}, we formulated two-dimensional cores using a different notation.
However, they were not designed for any effective treatment of signal boundaries.
In \cite{Barina2016,Barina2016b}, we have proposed DWT transform engine for JPEG 2000 encoders.
Also in this work we have not overcame the problem with effective treatment of signal boundaries.
Instead, we have introduced several buffers in order to circumvent the problem.

As it can be expected, we see a gap which can allow for significant simplifications and speedups.
Specifically, we propose a solution consisting of a mutable mapping of input coefficients onto output ones.

\section{Proposed Method}
\label{sec:proposed-method}

The section proposes a computation unit built using the lifting scheme technique.
The direct consequence of this formulation is the possibility of an elegant treatment of signal boundaries.
As mentioned above. the proposes unit is referred to as the core.

In this paragraph, some terminology necessary to understand the following text is clarified.
Lag $F$ describes a delay of the output samples with respect to the input samples.
The stage is used in the sense of the scheme step, usually the lifting step.
In linear algebra, such stage can be described by the linear operator (a matrix) mapping the input vector onto the output vector.

The following part of the section leads to the formulation of the core.
For demonstration purposes, only even-length signals are considered.
The single level of the discrete wavelet transform decomposes the input signal
\begin{align}
	\left( {a}^{0}_{n_{\vphantom{j}0}} \right)_{0 \le n < N}
\end{align}
of size $N$ samples into the resulting wavelet bands
\begin{align}
	\left( {d}^{1}_{n_{\vphantom{j}1}} \right)_{0 \le n_1 < N/2}, &&
	\left( {a}^{1}_{n_{\vphantom{j}1}} \right)_{0 \le n_1 < N/2}.
\end{align}

As a next step, a unit which continuously consumes the input signal ${a}^{0}$ and produces the output ${a}^{1},{d}^{1}$ subbands is proposed.
As mentioned above, this unit is referred to as the "core" in this paper.
As a consequence of the DWT nature, the core has to consume pairs of input samples.
The input signal is processed progressively from the beginning to the end, therefore in a single loop.
The corresponding output samples are produced with a lag $F$ samples depending on the underlying computation scheme.
The core requires access to an auxiliary buffer $ {{B}}. $
This buffers hold intermediate results of the underlying computation scheme.
The size of the buffer can be expressed as $\kappa$ coefficients, where $\kappa$ is the number of values that have to be passed between adjacent cores.

To simplify relations, two functions are introduced given by
\begin{align}
	\label{eqn:ThetaOmega}
	\Theta(n) = n+F, \text{ and }
	\Omega(n) = \lceil n /2 \rceil.
\end{align}
The function $\Theta(n)$ maps core output coordinates onto core input coordinates with the lag $F$.
The function $\Omega(n)$ maps the coordinates at the input level onto coordinates at the output level with respect to the chosen coordinate system.
Note that the $\Omega(n)$ can be defined in many ways.
We chose the definition (\ref{eqn:ThetaOmega}) that is compatible with JPEG 2000 standard.

The core transforms the fragment ${I}_{n}$ of an input signal onto the fragment ${O}_{n}$ of an input signal
\begin{align}
	{I}_{n} &=
	\begin{pmatrix} &
		{a}^{0}_{\Theta(n  )} &
		{a}^{0}_{\Theta(n+1)} &
	\end{pmatrix}^T, \\
	{O}_{n} &=
	\begin{pmatrix} &
		{a}^{1}_{\Omega(n  )} &
		{d}^{1}_{\Omega(n+1)} &
	\end{pmatrix}^T,
\end{align}
while updating the auxiliary buffer.

\begin{table*}
	\centering
	\caption[Individual linear transformations inside the mutable core.]{
		Individual linear transformations inside the mutable CDF 5/3 core.
		Also inverse lifting steps ${T}^{-1}_{\alpha,\Theta(n)}, {S}^{-1}_{\beta,\Theta(n)}$ are shown.
		Changes are displayed in different color.
	}
	\def\arraystretch{1.6}
	\begin{tabular}{|c|c|c|c|c|c|c|}
		\hline
			$\Theta(n)$ &
			${T}^{-1}_{\alpha,\Theta(n)}$ &
			${S}^{-1}_{\beta,\Theta(n)}$ &
			$T_{\alpha,\Theta(n)}$ &
			$S_{\beta,\Theta(n)}$ \\
		\hline
			$0$   &
				$\begin{bmatrix}
					0 & 0 & 1 & 0 \\ 
					\alpha & 0 & \alpha & 1 \\ 
					1 & 0 & 0 & 0 \\ 
					0 & 1 & 0 & 0    
				\end{bmatrix}$ &
				$\begin{bmatrix}
					1 & 0 & 0 & 0 \\ 
					0 & 1 & 0 & 0 \\ 
					0 & \beta & 1 & \beta \\ 
					0 & 1 & 0 & 0    
				\end{bmatrix}$ &
				$\begin{bmatrix}
					0 & 0 & 1 & 0 \\ 
					\color{red} 0 & 0 & \color{red} 2\alpha & 1 \\ 
					1 & 0 & 0 & 0 \\ 
					0 & 1 & 0 & 0    
				\end{bmatrix}$ &
				$\begin{bmatrix}
					1 & 0 & 0 & 0 \\ 
					0 & 1 & 0 & 0 \\ 
					0 & \beta & 1 & \beta \\ 
					0 & 1 & 0 & 0    
				\end{bmatrix}$ \\[1em]
			$1$   &
				$\begin{bmatrix}
					0 & 0 & 1 & 0 \\ 
					\alpha & 0 & \alpha & 1 \\ 
					1 & 0 & 0 & 0 \\ 
					0 & 1 & 0 & 0    
				\end{bmatrix}$ &
				$\begin{bmatrix}
					1 & 0 & 0 & 0 \\ 
					0 & 1 & 0 & 0 \\ 
					0 & \color{red} 2\beta & 1 & \color{red} 0 \\ 
					0 & 1 & 0 & 0    
				\end{bmatrix}$ &
				$\begin{bmatrix}
					0 & 0 & 1 & 0 \\ 
					\alpha & 0 & \alpha & 1 \\ 
					1 & 0 & 0 & 0 \\ 
					0 & 1 & 0 & 0    
				\end{bmatrix}$ &
				$\begin{bmatrix}
					1 & 0 & 0 & 0 \\ 
					0 & 1 & 0 & 0 \\ 
					0 & \beta & 1 & \beta \\ 
					0 & 1 & 0 & 0    
				\end{bmatrix}$ \\[1em]
			$\Theta(n)$   &
				$\begin{bmatrix}
					0 & 0 & 1 & 0 \\ 
					\alpha & 0 & \alpha & 1 \\ 
					1 & 0 & 0 & 0 \\ 
					0 & 1 & 0 & 0    
				\end{bmatrix}$ &
				$\begin{bmatrix}
					1 & 0 & 0 & 0 \\ 
					0 & 1 & 0 & 0 \\ 
					0 & \beta & 1 & \beta \\ 
					0 & 1 & 0 & 0    
				\end{bmatrix}$ &
				$\begin{bmatrix}
					0 & 0 & 1 & 0 \\ 
					\alpha & 0 & \alpha & 1 \\ 
					1 & 0 & 0 & 0 \\ 
					0 & 1 & 0 & 0    
				\end{bmatrix}$ &
				$\begin{bmatrix}
					1 & 0 & 0 & 0 \\ 
					0 & 1 & 0 & 0 \\ 
					0 & \beta & 1 & \beta \\ 
					0 & 1 & 0 & 0    
				\end{bmatrix}$ \\[1em]
			$\Theta(N-2)$ &
				$\begin{bmatrix}
					0 & 0 & 1 & 0 \\ 
					\alpha & 0 & \alpha & 1 \\ 
					1 & 0 & 0 & 0 \\ 
					0 & 1 & 0 & 0    
				\end{bmatrix}$ &
				$\begin{bmatrix}
					1 & 0 & 0 & 0 \\ 
					0 & 1 & 0 & 0 \\ 
					0 & \beta & 1 & \beta \\ 
					0 & 1 & 0 & 0    
				\end{bmatrix}$ &
				$\begin{bmatrix}
					0 & 0 & 1 & 0 \\ 
					\alpha & 0 & \alpha & 1 \\ 
					1 & 0 & 0 & 0 \\ 
					0 & 1 & 0 & 0    
				\end{bmatrix}$ &
				$\begin{bmatrix}
					1 & 0 & 0 & 0 \\ 
					0 & 1 & 0 & 0 \\ 
					0 & \beta & 1 & \beta \\ 
					0 & 1 & 0 & 0    
				\end{bmatrix}$ \\[1em]
			$\Theta(N)$ &
				$\begin{bmatrix}
					0 & 0 & 1 & 0 \\ 
					\color{red} 2\alpha & 0 & \color{red} 0 & 1 \\ 
					1 & 0 & 0 & 0 \\ 
					0 & 1 & 0 & 0    
				\end{bmatrix}$ &
				$\begin{bmatrix}
					1 & 0 & 0 & 0 \\ 
					0 & 1 & 0 & 0 \\ 
					0 & \beta & 1 & \beta \\ 
					0 & 1 & 0 & 0    
				\end{bmatrix}$ &
				$\begin{bmatrix}
					0 & 0 & 1 & 0 \\ 
					\alpha & 0 & \alpha & 1 \\ 
					1 & 0 & 0 & 0 \\ 
					0 & 1 & 0 & 0    
				\end{bmatrix}$ &
				$\begin{bmatrix}
					1 & 0 & 0 & 0 \\ 
					0 & 1 & 0 & 0 \\ 
					0 & \color{red} 0 & 1 & \color{red} 2\beta \\ 
					0 & 1 & 0 & 0    
				\end{bmatrix}$ \\
		\hline
	\end{tabular}
	\label{tab:mutable-cores}
\end{table*}

Finally, operations performed inside the core can be described using a matrix $C$ as the relationship
\begin{align}
	\label{eqn:core}
	\vect{y} = C \, \vect{x}
\end{align}
of the input vector
\begin{align}
	\vect{x} =
		{{B}} \concat
		{I}_{n}
\end{align}
with the output vector
\begin{align}
	\vect{y} =
		{{B}} \concat
		{O}_{n},
\end{align}
where $\concat$ denotes the concatenation operator.
The (\ref{eqn:core}) is the most fundamental equation of this thesis.
In this linear mapping, the matrix $C$ defines the core.

The meaning and the number of individual coefficients in ${{B}}$ is not firmly given.
The choice of the matrix $C$ involves a degree of freedom of the presented framework.
Some notes regarding this choice can be found in \Section{sec:discussion}.

\pagebreak
To keep the total number of wavelet coefficients equal to the number of input samples, symmetric border extension is widely used.
A particular variant of this extension is employed in JPEG 2000 standard.
Please, consult particular details with~\cite{Taubman2002}.

This paper describes the core which calculates the CDF 5/3 transform.
For the shortest possible lag $F=1$, it is easy to ensemble the core from (\ref{eqn:polyphase-cdf53}) as
\begin{align}
	C =
	\begin{bmatrix}
		1 & 0 & 0 & 0 \\ 
		0 & 1 & 0 & 0 \\ 
		0 & \beta & 1 & \beta \\ 
		0 & 1 & 0 & 0    
	\end{bmatrix}
	\begin{bmatrix}
		0 & 0 & 1 & 0 \\ 
		\alpha & 0 & \alpha & 1 \\ 
		1 & 0 & 0 & 0 \\ 
		0 & 1 & 0 & 0    
	\end{bmatrix},
\end{align}
for
\newcommand\wA[1]{\makebox[2.5em][l]{\scriptsize{}$\mathstrut{}#1$}}
\newcommand\wB[1]{\makebox[1.5em][l]{\scriptsize{}$\mathstrut{}#1$}}
\begin{align}
	\vect{x} &= \begin{bmatrix} & B_a & B_d & a^0_{\wA{n+1}}   & a^0_{\wB{n}} & \end{bmatrix}^T \text{and}\\
	\vect{y} &= \begin{bmatrix} & B_a & B_d & a^1_{\wA{n/2-1}} & d^1_{\wB{n/2}} & \end{bmatrix}^T.
\end{align}
The $B$ comprises $\kappa=2$ elements.
For the purposes of the discussion, only even-length signals are considered.
The core consists of two stages suitable for hardware pipelining.

As mentioned earlier, the core processes the signal in the single loop.
The naive way of border handling is described first.
Due to the symmetric extension, the core begins the processing at a certain position before the start of the actual signal sequence.
Similarly, the processing stops at a certain position after the end of the signal.
The samples outside the actual signal are mirrored into the valid signal area.
This processing introduces the need for buffering of the input at least at the beginning and the end of the signal.
Such buffering breaks the ability of simple stream processing, especially considering the multi-scale decomposition.
All approaches referenced in \Section{sec:related-work} also suffer from this issue.

\begin{figure*}
	\centering
	\def\svgwidth{.75\linewidth}
	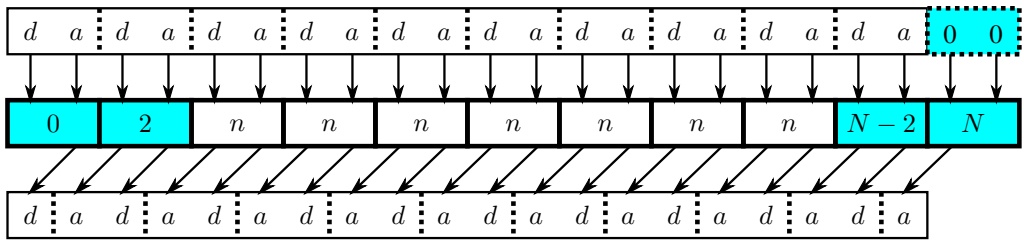
	\caption[Signal processing using the mutable core.]{
		Signal processing using the mutable core.
		The input position on the original location $n$ is shown.
		The first and last two cores (highlighted) differ from the others.
	}
	\label{fig:mutable-cores}
\end{figure*}

The situation can be neatly resolved changing the core near the signal border.
In more detail, the "mutable" core performs 5 different calculations depending on the position in relation to the input signal.
Therefore, the core comprises $2\times5$ slightly different steps (stages) in total.
This can be written in matrix notation as
\begin{align}
	\vect{y} = S_{\beta,\Theta(n)} \, T_{\alpha,\Theta(n)} \, \vect{x},
\end{align}
where $T_{\alpha,\Theta(n)}, S_{\beta,\Theta(n)}$ are the linear transformations of the predict and update stages performed at the subsampled output position $\Theta(n)$.
These coefficients are generated in $T_{\alpha,\Theta(n)}$ so that these can be used by $T_{\alpha,\Theta(n+2)}$ at the same time at which $S_{\beta,\Theta(n+2)}$ runs.
It is essential that the coefficients $B_a, B_d$ are initially set to zero.
The output signal is generated with the lag $F=1$ sample with respect to the input signal.
The input $a$ samples outside of the input signal are treated as zeros.
Similarly, the output $a, d$ coefficients outside of the output signal are discarded.
The following table describes the individual $T_{\alpha,\Theta(n)}, S_{\beta,\Theta(n)}$ transformations.
The transform is defined using the $\alpha, \beta$ constants.
\Table{tab:mutable-cores} enumerates the individual core stages.
In addition, \Figure{fig:mutable-cores} illustrates their usage.
The very first and very last cores access outside the signal.
The input samples already split into $a, d$ subbands.
As a result, the signal is transformed without buffering, possibly on a multi-scale basis.

\vspace{\baselineskip}
\section{Discussion}
\label{sec:discussion}

So far, only the CDF 5/3 wavelet has been discussed as an illustrative example.
However, the presented computation scheme is general.

One can identify the core in arbitrary underlying lifting schemes.
However, its implementation can be obscure in some cases mainly due to the increasing number of intermediate results in auxiliary buffers.
For his reason, a lifting factorization employing steps in the form of degree-1 filters is a proven choice.
Many such factorizations of various wavelets have been presented in the literature, e.g. \cite{Daubechies1998}.
Moreover, the symmetric filters with lengths $2 x \pm 1$, as is case of the CDF 5/3 and 9/7 filters, can be implemented through some sequence of lifting steps having this particular form.
See \cite{Taubman2002} for details.

In parallel environments, the presented scheme can be "unwrapped" to fit the number of computing units (threads).
Instead of passing the intermediate results through the buffer $B$, an direct exchange of them then arises.
Unfortunately, this exchange introduces the need for a synchronization barrier, that can be a bottleneck of the processing.

The multi-scale decomposition can be easily constructed by chaining the cores into a series.
In this case, only the $a$ coefficients are linked to the next decomposition level.
Each subsequent decomposition level causes additional delay $F$ at the half sampling rate.

Considering multi-dimensional signals, multi-dimensional cores can be constructed as the tensor product of \mbox{1-D} cores.
For example, the \mbox{2-D} core consumes the $2\times2$ fragment of the input signal $a^0$ and immediately produces the four-tuple $(a^1,h^1,v^1,d^1)$ of resulting subbands.

The core $C$ can also be internally reorganized in order to minimize some of the resources.
In \cite{Barina2015c}, we demonstrated this property on FPGA where the minimization of the core latency had a direct impact on the utilization of flip-flop (FF) circuits and look-up tables (LUT).

The construction of the matrices themselves is governed by simple rule.
When the coefficient outside a valid signal sequence would be accessed, the zero is put into the matrix at the corresponding place.
Additionally, the symmetrization is obtained by placing the factor of 2 at desired locations at the same moment.
In this paper, we have discussed only the symmetric signal extension.
Nevertheless, the matrices for any other extension could be constructed.
For zero-padding, the factor of 2 should be eliminated everywhere in \Table{tab:mutable-cores}.

One can simply verify the correctness of the scheme presented by its unwrapping.
The data-flow graph obtained shows the mirroring at the beginning and end of the signal.

In conclusion, this paper presents nice idea to take of boundary effects in DWT processing.
Usually some form of wrap-around or replication is necessary to take care of the image boundaries.
But we have proposed a slightly different way of handling this which then makes efficient parallel or pipelined processing possible.

\vspace{\baselineskip}
\section{Conclusions}
\label{sec:conclusions}

We have focuses on treatment of signal boundaries for computing of the discrete wavelet transform.
In the state-of-the-art methods, the treatment of signal boundaries is performed in a complicated way.
This way causes buffering of the signal at least at the beginning and end of data.

In this paper, we have overcome this issue.
This was accomplished using the proposed mutable core which performs the transform in a single loop.
Using this core, the complete transform can be evaluated as a running scheme.
In other words, no buffering is required anywhere.
The resulting coefficients are exactly the same as with the original lifting scheme.

The future work we would like to do includes
an extension to more complicated lifting factorizations,
and a hardware implementation of the core proposed.

\vspace{\baselineskip}
\section*{Acknowledgement}
This work was supported by
the TACR Competence Centres project V3C -- Visual Computing Competence Center (no. TE01020415), and
the Ministry of Education, Youth and Sports of the Czech Republic from the National Programme of Sustainability (NPU II) project IT4Innovations excellence in science -- LQ1602.

\newcommand{\BIBdecl}{\setlength{\itemsep}{.5 em}}
\bibliographystyle{IEEEtranDoi}
\bibliography{sources,publications}

\end{document}